\documentclass[oneside,a4paper,11pt,reqno]{amsart}

\usepackage[margin=35mm]{geometry}
\usepackage[utf8]{inputenc}
\usepackage{mathrsfs,dsfont,hyperref,amsmath,amssymb,amsthm,amsfonts,amstext,amsopn,amsxtra,mathrsfs,mathtools}
\usepackage[capitalize]{cleveref}
\usepackage{esint,enumitem,bookmark}
\usepackage{braket}
\usepackage{stmaryrd}

\newtheorem{lemma}{Lemma}
\newtheorem{theorem}[lemma]{Theorem}
\newtheorem{conjecture}[lemma]{Conjecture}
\newcommand{\R}{\mathbb{R}}
\newcommand{\C}{\mathbb{C}}
\newcommand\1{{\ensuremath {\mathds 1} }}
\newcommand{\cM}{\mathcal{M}}
\newcommand{\cE}{\mathcal{E}}
\newcommand{\cW}{\mathcal{W}}
\newcommand{\cL}{\mathcal{L}}
\newcommand{\tr}{{\rm Tr}\,}
\newcommand{\supp}{{\rm Supp}}
\renewcommand{\geq}{\geqslant}
\renewcommand{\leq}{\leqslant}
\newcommand{\eps}{\varepsilon}

\newcommand{\bx}{\mathbf{x}}
\newcommand{\ad}{\mathrm{ad}}
\def\d{\,{\rm d}}

\usepackage{hyperref}

\title[Dilute Bose gas with 3-body interaction]{Dilute Bose gas with three-body interaction:\\recent results and open questions}

\author[P.T. Nam]{Phan Th\`anh Nam}
	\address{Department of Mathematics, LMU Munich, Theresienstrasse 39, 80333 Munich, and Munich Center for Quantum Science and Technology, Schellingstr. 4, 80799 Munich, Germany} 
	\email{nam@math.lmu.de}

\author[J. Ricaud]{Julien Ricaud}
	\address{Centre de Math\'{e}matiques Appliqu{\'e}es, {\'E}cole polytechnique, 91128 Pa\-lai\-seau Cedex, France}
	\email{julien.ricaud@polytechnique.edu}

\author[A. Triay]{Arnaud Triay}
	\address{Department of Mathematics, LMU Munich, Theresienstrasse 39, 80333 Munich, and Munich Center for Quantum Science and Technology, Schellingstr. 4, 80799 Munich, Germany}
	\email{triay@math.lmu.de}

\begin{document}

\begin{abstract}
	We review our recent study on the ground state energy of dilute Bose gases with three\nobreakdash-body interactions. The main feature of our results is the emergence of the 3D energy-critical Schr{\"o}\-dinger equation to describe the ground state energy of a Bose--Einstein condensate, where the nonlinearity strength is determined by a zero scattering problem. Several open questions are also discussed.
\end{abstract}

\maketitle

%%%%%%%%%%%%%%%%%%%%%%%%%%%%%%%%%%%%%%%%%%%%%%%%%%%%%%%%% 
\section{Introduction} 
%%%%%%%%%%%%%%%%%%%%%%%%%%%%%%%%%%%%%%%%%%%%%%%%%%%%%%%%%
Bose--Einstein condensation (BEC) is a fantastic playground for both probing the principles of quantum mechanics and exploring novel physics. Following the first experimental realizations of BEC by Cornell, Wieman and Ketterle in 1995~\cite{CorWie-95,Ketterle-95}, there has been a regain of interest in the mathematical physics community for rigorous results starting from first principles. It began with the proof of the correct lower bound for the ground state energy in the dilute limit, by Lieb and Yngvason in 1998~\cite{LieYng-98}, which complemented the upper bound proved by Dyson in 1957~\cite{Dyson-57}. Then followed the derivation of the Gross--Pitaevskii functional by Lieb, Yngvason and Seiringer in 2000~\cite{LieSeiYng-00}, as well as the proof of BEC for the ground state in this regime by Lieb and Seiringer in 2002~\cite{LieSei-02}. Those works deal with the case of a two\nobreakdash-body interaction, which is the typical setting for a dilute gas where the probability of having more than two particles in the same neighborhood is very low. However, in some particular settings, the effects of many-body interactions can contribute to the leading order. In this review, we are interested in the appearance of three\nobreakdash-body interactions, particularly focusing on the generalization to this case of the results in~\cite{Dyson-57,LieYng-98,LieSeiYng-00,LieSei-02}. 

Although three\nobreakdash-body interactions are often considered to be lower order corrections to two\nobreakdash-body interactions, in some cases they can be artificially enhanced by internal coupling up to becoming prominent~\cite{Petrov-14,Hammond-21}, in a similar manner to that the two\nobreakdash-body scattering length can be tuned via Feshbach resonance. Note that in the physics literature, three\nobreakdash-body interactions are sometimes taken into account as non-conservative forces that reduce the number of particles in the trap over time via a phenomenon called three\nobreakdash-body recombination: this is \emph{not} the subject of our work. Here we consider on the contrary conservative three\nobreakdash-body repulsive interactions.

\bigskip
\noindent 
{\bf Acknowledgments.} We thank the referee for many helpful suggestions. We received funding from the Deutsche Forschungsgemeinschaft (DFG, German Research Foundation) under Germany's Excellence Strategy (EXC-2111-390814868). J.R. also acknowledges financial support from the French Agence Nationale de la Recherche (ANR) under Grant No. ANR-19-CE46-0007 (project ICCI).

%%%%%%%%%%%%%%%%%%%%%%%%%%%%%%%%%%%%%%%%%%%%%%%%%%%%%%%%%
\section{Three-body interactions and scattering energy}
%%%%%%%%%%%%%%%%%%%%%%%%%%%%%%%%%%%%%%%%%%%%%%%%%%%%%%%%%
We model interactions between three particles via a potential which depends only on relative coordinates:
\begin{equation} \label{eq:U-V}
	U(x_1,x_2,x_3) = V(x_1-x_2,x_1-x_3)\,,
\end{equation}
where $V: \R^3 \times \R^3 \to \R$ is nonnegative, bounded, and compactly supported. In order to preserve the bosonic symmetry of the particles, $U$ should be invariant under permutations of the three variables. This translates into the following \emph{three\nobreakdash-body symmetry} of $V$ 
\begin{equation}\label{eq:sym}
	V(x,y) = V(y,x) \quad \textrm{ and } \quad V(x-y,x-z) = V(y-x,y-z) = V(z-y,z-x)\,.
\end{equation}
As we will explain below, this symmetry plays a crucial role in the zero-scattering energy of three bosons. 

%%%%%%%%%%%%%%%%%%%%%%%%%%%%%%
\subsection{Scattering energy}
%%%%%%%%%%%%%%%%%%%%%%%%%%%%%%
In any dimension $d\geq 3$, we can define the {\em zero-scattering energy} of a compactly supported potential $0\le v \in L^\infty(\R^d)$ by 
\begin{equation} \label{eq:def-scat-energy}
	b(v):= \inf_{\varphi \in \dot H^1(\R^d)} \int_{\R^d} \left( 2|\nabla \varphi( x)|^2 + v(x) |1-\varphi(x)|^2 \right) \d x\,,
\end{equation}
where $\dot H^1(\R^d)$ is the space of functions $g:\R^d \to \C$ vanishing at infinity and satisfying $|\nabla g|\in L^2(\R^d)$. This variational problem has a unique nonnegative optimizer $0\le \varphi \le 1$, and $f:= 1-\varphi$ satisfies 
\[
	b(v) = \int_{\R^{d}} v(x) f(x)\, \d x\,.
\]
In fact, the minimization problem~\eqref{eq:def-scat-energy} naturally generalizes to the case of hard sphere potentials
\[
	v_{\rm hs}(x) =
	\begin{cases}
		\infty, &\quad |x|<a\\
		0, &\quad |x|>a 
	\end{cases},
\]
for which we find $b(v_{\rm hs})=c_d a^{d-2}$. Thus $b(v)^{1/(d-2)}$ plays the role of the {\em scattering length}, up to a universal factor. 

On one hand, when $d = 3$ it is well-known that the scattering solution $\varphi(x)$ of~\eqref{eq:def-scat-energy} behaves like $a/|x|$ at infinity, where the constant $a=(8\pi)^{-1}b(v)>0$ is the scattering length of $v$. This is the typical situation associated with two\nobreakdash-body interactions studied in~\cite{LieYng-98,LieSeiYng-00,LieSei-02}. 

On the other hand, we are interested here in the case $d=6$, which is associated to three\nobreakdash-body interactions as in~\eqref{eq:U-V}. In this case, $b(v)^{1/4}$ plays the role of the scattering length.

%%%%%%%%%%%%%%%%%%%%%%%%%%%%%%
\subsection{Modified scattering energy.}
%%%%%%%%%%%%%%%%%%%%%%%%%%%%%%
Because of the choice of relative coordinates $U(x_1,x_2,x_3) = V(x_1-x_2,x_1-x_3)$, the scattering problem in $\R^9$ for $U$ is naturally associated to an effective scattering problem in $\R^6$ for $V$, which is however slightly different from~\eqref{eq:def-scat-energy} due to the removal of the center of mass. Let us explain this in detail. Considering the change of coordinates
\[
	r_{1} = \frac{1}{3}(x_1+x_2+x_3)\,, \quad r_2 = x_1 -x_2 \quad \textrm{ and } \quad r_3 =x_1-x_3\,,
\]
we can rewrite the three\nobreakdash-body scattering operator as 
\begin{multline}\label{eq:remove-center}
	-\Delta_{x_1} - \Delta_{x_2} - \Delta_{x_3} + V(x_1-x_2,x_1-x_3)\\
	\begin{aligned}[b]
		&= \left( \frac{1}{3} p_{r_1} + p_{r_2} + p_{r_3} \right)^2 + \left(\frac{1}{3} p_{r_1} - p_{r_2} \right)^2 + \left( \frac{1}{3} p_{r_1}-p_{r_3} \right)^2 + V(r_1,r_2) \\
		&= \frac{1}{3}p_{r_1}^2 + 2(p_{r_2}^2 + p_{r_3}^2 + p_{r_2} p_{r_3} ) + V(r_2,r_3) \\
		&= \frac{1}{3}p_{r_1}^2 + \left\langle \begin{pmatrix}
		p_{r_2} \\
		p_{r_3}
	\end{pmatrix}, \begin{pmatrix}
		2 & 1 \\
		1 & 2
	\end{pmatrix} \begin{pmatrix}
		p_{r_2} \\
		p_{r_3} 
	\end{pmatrix} \right\rangle + V(r_2,r_3) \,,
	\end{aligned}
\end{multline}
where $p_r= -{\bf i}\nabla_r$. Note that $p_{r_1}^2 \ge 0$ and since it is completely separated from the last two terms on the r.h.s.\ of~\eqref{eq:remove-center}, we can simply drop it when considering low-energy states. This results in the following operator on $L^2(\R^6)$: 
\begin{equation}\label{eq:intro-M-ope}
	\left\langle
	\begin{pmatrix}
		p_{x_1} \\
		p_{x_2}
	\end{pmatrix},
	\begin{pmatrix}
		2 & 1 \\
		1 & 2
	\end{pmatrix}
	\begin{pmatrix}
		p_{x_1} \\
		p_{x_2}
	\end{pmatrix}
	\right\rangle + V(x_1, x_2) = 2 | \cM \nabla_{\R^6} |^2 + V(x_1, x_2)\,,
\end{equation}
where the matrix $\cM: \R^3 \times \R^3 \to \R^3 \times \R^3$ is given by
\[
	\cM
	= \frac{1}{2\sqrt{2}}
	\begin{pmatrix}
		\sqrt{3}+1 & \sqrt{3}-1 \\
		\sqrt{3}-1 & \sqrt{3}+1 
	\end{pmatrix}
	=\left( \frac{1}{2}
	\begin{pmatrix}
		2 & 1 \\
		1 & 2
	\end{pmatrix}
	\right)^{1/2}.
\]
Thus the effective scattering energy associated with~\eqref{eq:intro-M-ope} is 
\begin{equation}\label{eq:def_b}
	b_{\cM}(V):= \inf_{\varphi \in \dot H^1(\R^6)} \int_{\R^6} \left( 2| \cM \nabla_\bx \varphi(\bx)|^2 + V(\bx) |1-\varphi(\bx)|^2 \right) \d \bx\,.
\end{equation}
By a change of variables, this value can be related to the definition in ~\eqref{eq:def-scat-energy} through
\[
	b_{\cM}(V) = b(V(\cM \cdot)) \det \cM \,.
\]
As proved in~\cite{NamRicTri-21}, the variational problem~\eqref{eq:def_b} has an optimizer $\omega_{\cM}=1-f_{\cM}$ where $f_{\cM}:\R^6\to \R$ satisfies the three\nobreakdash-body symmetry~\eqref{eq:sym} and solves the scattering equation
\begin{equation}\label{eq:scat_M}
	-2|\cM \nabla_\bx|^2 f_{\cM} (\bx) + V(\bx) f_{\cM}(\bx) =0, \quad \forall \, \bx\in \R^6\,, \quad \textrm{ and } \quad \lim_{|\bx|\to \infty} f_{\cM}(\bx)=1\,.
\end{equation}
Then the modified scattering energy can be written as 
\[
	b_{\cM}(V) = \int_{\R^6} V(\bx) f_{\cM}(\bx)\, \d \bx \,.
\]
Since $b_{\cM}(V)$ and $b(V)$ have the same order of magnitude, $b_{\cM}(V)^{1/4}$ plays the role of the scattering length.

%%%%%%%%%%%%%%%%%%%%%%%%%%%%%%%%%%%%%%%%%%%%%%%%%%%%%%%%% 
\section{The thermodynamic limit} \label{sec:TL}
%%%%%%%%%%%%%%%%%%%%%%%%%%%%%%%%%%%%%%%%%%%%%%%%%%%%%%%%%
Let us consider $N$ bosons in $\Omega = [-L/2,L/2]^3$, for some $L>0$, interacting via a three\nobreakdash-body potential $V: \R^3 \times \R^3 \to [0,\infty)$. The system is described by the Hamiltonian
\begin{equation}\label{eq:H_NL}
	H_{N,L} = \sum_{i=1}^N -\Delta_{x_i} + \sum_{1\le i < j < k \le N} V(x_i - x_j,x_i-x_k)
\end{equation}
acting on the bosonic space $L^2_s(\Omega^N)$, and where $-\Delta$ denotes the Laplacian with Neumann boundary conditions. Since $V$ satisfies~\eqref{eq:sym}, the Hamiltonian $H_N$ leaves $L^2_s(\Omega^N)$ invariant.

Our main result in~\cite{NamRicTri-22} concerns the thermodynamic ground state energy per unit volume which is defined for any $\rho > 0$ by
\begin{equation}\label{eq:erho}
	e_{\rm 3B}(\rho) := \lim_{\substack{N \to \infty \\ N / |\Omega| \to \rho}} \inf_{\|\Psi\|_{L^2\left(\Omega^N\right)}^2 = 1} \frac{\langle \Psi, H_{N,L} \Psi \rangle}{|\Omega|}\,.
\end{equation}
This limit exists and does not depend on the boundary conditions, nor on the choice of $\Omega \to \R^3$ when $N\to\infty$ as long as it is regular enough~\cite{Ruelle}.

In~\cite{NamRicTri-22} we estimated $e_{\rm 3B}(\rho)$, in the dilute limit, in terms of the scattering energy of the interaction potential $V$ and proved the following result.

\begin{theorem}[Ground state energy in the low density regime]\label{thm_GS_energy_low_density}
	Let $0\le V \in L^\infty(\R^6)$ be compactly supported and satisfy the three\nobreakdash-body symmetry~\eqref{eq:sym}. Then, in the dilute limit $Y:=\rho b_{\cM} (V)^{3/4} \to 0$, the thermodynamic ground state energy per unit volume in~\eqref{eq:erho} satisfies
	\[
		e_{\rm 3B}(\rho) = \frac{1}{6}b_{\cM}(V) \rho^3 (1 + \mathcal O(Y^{\nu}))
	\]
	for some constant $\nu>0$. 
\end{theorem}

The reader may think of $V$ as a fixed potential and interpret Theorem~\ref{thm_GS_energy_low_density} as a low-density result ($\rho\to 0$). However, it is helpful to keep in mind that the dimensionless parameter $Y=\rho b_{\cM} (V)^{3/4}$ is the right quantity to measure the diluteness of the system: the condition $Y\to 0$ tells us that the scattering length of the interaction ($\sim b_{\cM} (V)^{1/4}$) is much smaller than the average distance between particles ($\sim \rho^{-1/3}$). Hence, Theorem~\ref{thm_GS_energy_low_density} can in principle be also used to predict the ground state energy in different situations when both $\rho$ and $V$ vary. In particular, this will be consistent with the Gross--Pitaevskii regime discussed later. 

As explained in~\cite[Remark 2]{NamRicTri-22}, when both $\rho$ and $V$ vary, the error $\mathcal O(Y^{\nu})$ in Theorem~\ref{thm_GS_energy_low_density} is uniform in $V$ if $R_0/b_{\cM}(V)^{1/4}$, $\|V\|_{L^1} b_{\cM}(V)^{-1}$ and $\|V\|_{L^\infty} b_{\cM}(V)^{1/2}$ are uniformly bounded, where $R_0$ is the range of $V$. Actually, the lower bound holds uniformly as soon as $R_0$ is bounded, hence it is easily extendable to hard-core potentials. On the other hand, our estimate of the upper bound depends on both $\|V\|_{L^1}$ and $\|V\|_{L^\infty}$, making hard-core potentials not accessible by our proof, even though we believe the result to hold for these potentials too.

In the case of two\nobreakdash-body interactions, the leading order of the ground state energy was proved by Dyson~\cite{Dyson-57} (upper bound) and Lieb--Yngvason~\cite{LieYng-98} (lower bound). The proof of Theorem~\ref{thm_GS_energy_low_density} is more difficult than that of the two\nobreakdash-body interaction case, although several ideas from~\cite{Dyson-57} and~\cite{LieYng-98} are still very helpful. For the lower bound, we will need to introduce a new {\em Dyson lemma}, because the existing tools in~\cite{LieYng-98} do not apply directly to potentials that are not spherically symmetric, like the three\nobreakdash-body potential. For the upper bound, we are not able to adapt the trial state from~\cite{Dyson-57}. Instead we will introduce a unitary transformation in the spirit of {\em Bogoliubov's approximation} but where the relevant correlation kernel is related to a cubic creation operator $a^\dagger a^\dagger a^\dagger$ instead of a quadratic one $a^\dagger a^\dagger$. The main ingredients of the proof are explained below. 

%%%%%%%%%%%%%%%%%%%%%%%%%%%%%%
\subsection*{Sketch of the proof: reduction to smaller boxes}
%%%%%%%%%%%%%%%%%%%%%%%%%%%%%%
For both the lower and upper bounds, we divide $\Omega = [-L/2,L/2]^3$ into smaller boxes of side length $\ell >0$. By dilation, we are led to consider the Hamiltonian
\[
	\widetilde{H}_{n,\ell} = \sum_{i=1}^n -\Delta_{x_i} + \sum_{1\le i < j < k \le n} \ell^2 V\left(\ell(x_i-x_j,x_i-x_k)\right)\quad \text{ on }L^2_s\!\left([-1/2,1/2]^{3n}\right),
\]
where $n$ stands for the number of particles in the box $[-1/2,1/2]^3$. Indeed, denoting $\mathcal U \Psi = \ell^{3n/2}\Psi(\ell \cdot)$, one has $ H_{n,\ell} = \ell^{-2} \mathcal U^* \widetilde{H}_{n,\ell} \mathcal U$, which acts on $L^2_s([-\ell/2,\ell/2]^{3n})$, and where we recall that $H_{n,\ell}$ is the Hamiltonian defined in~\eqref{eq:H_NL}.

One important length scale is the Gross--Pitaevskii (GP) one, for which the gap of the kinetic operator is of the same order as the ground state energy per particle
\[
	1 \simeq n^2 b_{\cM}(\ell_{\rm GP}^{2}V(\ell_{\rm GP} \cdot)) \quad \iff \quad \ell_{\rm GP} \simeq \frac{1}{\rho b_{\cM}(V)^{1/2}}\,,
\]
where we used that $n\simeq \rho \ell^3$ and the scaling property $b_{\cM}(\ell^{2}V(\ell \cdot)) = \ell^{-4}b_{\cM}(V)$. By analogy with the two\nobreakdash-body case, we can define an effective scattering length $a = b_{\cM}(V)^{1/4}$, so that $\ell_{\rm GP} \simeq a / (\rho a^3)$, where we recall that $\rho a^3$ is the dimensionless diluteness parameter.

%%%%%%%%%%%%%%%%%%
\subsubsection*{Lower bound}
%%%%%%%%%%%%%%%%%%
The overall strategy of the proof of the lower bound in Theorem~\ref{thm_GS_energy_low_density} follows the one in~\cite{LieYng-98}. By non-negativity of the potential we can discard the interaction between boxes without increasing the energy, and we use a convexity argument to control the number of particles in the boxes. We choose a length scale much shorter than the Gross--Pitaevskii length scale $\ell \ll \ell_{\rm GP}$, for which the gap of the kinetic operator is large so that we can treat the interaction potential as a perturbation and apply the Temple inequality. However, we cannot do so directly, we first need to renormalize the interaction potential using a version of Dyson's lemma adapted to the modified scattering problem~\eqref{eq:def_b}.
\begin{lemma}[Dyson's lemma for non-radial potentials]\label{lem:dyson_lemma}
	Let $R_2 / 2 > R_1 > R_0 > 0$, $d\ge 3$, $\Omega$ be an open set with $\{|\bx| \le R_2\} \subset \Omega \subset \R^{d}$, and $0\le v \in L^\infty(\R^d)$ with $\supp\, v \subset \{ |\bx| \le R_0\}$. Then there exists $0\le U \in C(\R^d)$ with $\int_{\R^d} U = 1$ and $\supp \, U \subset \{ R_1\le |\bx| \le R_2\}$ such that the following operator inequality holds
	\[
		- 2 \cM \nabla_\bx \1_{\{|\bx| \le R_2\}} \cM \nabla_\bx + v(\bx) \ge b_{\cM}(v) \left( 1- \frac{C_d R_0}{R_1}\right) U (\bx) \quad \text{ on }L^2(\Omega)\,,
	\]
	with a constant $C_d>0$ depending only on the dimension $d$. 
\end{lemma}
Note that, compared to the usual Dyson lemma as in~\cite{LieSeiSolYng-05}, we cannot use the radial assumption of the potential (except for the trivial potential) as it would contradict the three\nobreakdash-body symmetry~\eqref{eq:sym}. Lemma~\ref{lem:dyson_lemma} allows us to sacrifice part of the kinetic energy in order to bound below the singular potential $\ell^{2} V(\ell \cdot)$ by a softer one $R^{-6} U(R^{-1}\cdot)$, for some $R \gg \ell^{-1}$, which satisfies $\int_{\R^6} U = 1$. Using this lemma, we can essentially bound below $\widetilde{H}_{n,\ell}$ by
\[
	\eps \sum_{i=1}^n -\Delta_{x_i} + \ell^{-4} b_{\cM}(V) \sum_{1\le i < j < k \le n} R^{-6} U\left(R^{-1}(x_i-x_j,x_i-x_k)\right),
\]
where we kept $\eps$ of the kinetic energy. 

Then, following~\cite{LieYng-98} again, we use the Temple inequality for the soft potential $R^{-6} U(R^{-1}\cdot)$. To that purpose, we need the gap of the kinetic operator to dominate the expectation of the interaction potential against the constant function:
\[
	\eps \gtrsim n^3 \ell^{-4} b_{\cM}(V)\,,
\]
which is made possible by the choice $\ell \ll \ell_{\rm GP}$. The Temple inequality gives
\begin{equation}\label{eq:bound_H_nl}
	\inf \sigma(\widetilde{H}_{n,\ell}) \geq \frac{1}{6}n^3 \ell^{-4} b_{\cM}(V) \left(1 + \mathcal O\!\left(\left(\rho^3 b_{\cM}(V)\right)^\nu\right)\right),
\end{equation}
for some $\nu > 0$. This is the energy in one box and we have to multiply it by the number of boxes $(L/\ell)^3$. Recalling that $n \simeq \rho \ell^3$, and that $H_{n,\ell} = \ell^{-2} \mathcal U \widetilde{H}_{n,\ell} \mathcal U^*$, we obtain the claim.

%%%%%%%%%%%%%%%%%%
\subsubsection*{Upper bound}
%%%%%%%%%%%%%%%%%%
For the upper bound, we are not able to adapt Dyson's analysis in~\cite{Dyson-57} because the estimate of four\nobreakdash-body contributions based on the ``nearest neighbor technique" becomes very complicated. Instead, we follow another approach which requires more regularity conditions on the interaction potential. In particular we cannot treat hard-core potentials as in~\cite{Dyson-57}. In this approach, we first localize particles in smaller boxes of side length $\ell \ll L$ in order to improve the spectral gap of the kinetic operator. Then, in each box, we rely on the computations made in~\cite{NamRicTri-21} where the trapped case in the Gross--Pitaevskii regime was dealt with. The trapped case corresponds essentially to the Dirichlet boundary conditions, to which the computations in~\cite{NamRicTri-21} extend easily. Heuristically, the price to pay for localizing $N$ particles in boxes of side length $\ell$ is $N \ell^{-2}$, and since we want this error to be subleading in the large volume limit, we require
\[
	\frac{N}{L^3} \ell^{-2} \ll \rho^3 b_{\cM}(V) \iff \ell \gg \ell_{\rm GP}\,.
\]
Therefore, we need to consider boxes of side length much larger than the Gross--Pitaevskii length scale in order to be able to ignore the dependence on the boundary conditions.
Indeed, we can take 
\[
	\ell = \ell_{\rm GP} Y^{-\alpha}
\]
where $Y=\rho b_{\cM}(V)^{3/4}\to 0$ is the dimensionless diluteness parameter and $\alpha>0$ is a small constant. The main point of the analysis is to show that, at this length scale, the leading order energy remains essentially the same as that of the Gross--Pitaevskii regime.

Let us briefly explain our choice of ansatz. Since it is easier to work directly in the grand-canonical ensemble and because we are only interested in the leading order of the energy, by the equivalence of ensembles, it is enough to consider
\begin{equation}\label{eq:H_n_fock}
	\mathbb{H}_{\ell} = \int_{\R^3} \nabla_x a^\dagger_x \nabla_x a_x + \frac{1}{6}\int_{\R^9} \ell^2 V(\ell(x-y,x-z)) a^\dagger_x a^\dagger_y a^\dagger_z a_x a_y a_z
\end{equation}
acting on the Fock space 
\[
	\mathcal F_s \!\left(L^2 \left([-1/2,1/2]^3 \right) \right) = \bigoplus_{n\geq 0} L^2_s \!\left( [-1/2,1/2]^{3n} \right)\,.
\]
We easily check that $\mathbb{H}_{\ell} = \bigoplus_{n\geq 0} \widetilde{H}_{n,\ell}$. Let us denote $u_0 \equiv 1$ the condensate wave function and $P = 1-Q$ the orthogonal projection onto it. Rigorously, we cannot take $u_0 \equiv 1$ because it does not satisfy Dirichlet boundary conditions. However, as explained earlier, the length scale has been chosen large enough so that the localization error is subleading and we make this abuse for clarity. Bogoliubov's original approximation was to factor out the condensate by implementing the c\nobreakdash-number substitution 
\[
	a^\dagger_x \simeq \int_{\R^3} Q(x,y) a^\dagger_y + \sqrt n =: c^\dagger_x + \sqrt n\,.
\]
The reason behind this heuristics is that one can expect that 
\[
	\int_{\R^3} P(x,y) a^\dagger_y = a^\dagger(u_0) \simeq \sqrt{a^\dagger (u_0) a(u_0)} \simeq \sqrt n\,,
\]
if most of the particles are in the condensed state. Since we work in the grand-canonical picture, we can rigorously implement this using the Weyl transform
\[
	\cW = \exp\left(\sqrt n a^\dagger(u_0) - \sqrt{n} a(u_0)\right),
\]
whose action on the creation and annihilation operators is given by
\[
	\cW^* a^\dagger_x \cW = a^\dagger_x + \sqrt n \quad \textrm{ and } \quad \cW^* a_x \cW = a_x + \sqrt n\,.
\]
Inserting these relations into~\eqref{eq:H_n_fock} and expanding, we find that $\cW^* \mathbb{H}_{\ell}\cW$ is the sum of terms having $0,1,\dots,5$ or $6$ occurrences of the creation and annihilation operators:
\begin{equation}\label{eq:excitation_H}
	\cW^* \mathbb{H}_{\ell}\cW = \d\Gamma(-\Delta) + \sum_{i=0}^6 \cL_i\,.
\end{equation}
The term $\cL_0$ is a mean-field term and contributes to the leading order in the energy:
\[
	\cL_0 = \frac{1}{6} n^3 \ell^{-4} \int_{\R^6} V\,.
\]
We already know that $\cL_0$ is not a good approximation of the ground state energy. Indeed, by taking $\varphi = 0$ in the minimization problem~\eqref{eq:def_b}, we obtain $\int_{\R^6} V > b_{\cM}(V)$. It turns out that the remaining part of the energy is created by the cubic term
\begin{equation}\label{eq:L3}
	\cL_3 \simeq \int_{\R^9} n^{3/2} \ell^2 V(\ell(x-y,x-z)) (a^\dagger_x a^\dagger_y a^\dagger_y + a_x a_y a_z)\,.
\end{equation}
In order to extract the energy created by $\cL_3$, we conjugate the Hamiltonian by a unitary transform $\exp(B)$, where the operator $B$ is skew-adjoint and cubic in creation and annihilation operators:
\[
	B := -\frac{1}{6}\int_{\R^9} n^{3/2} \omega_{\cM} (\ell(x-y,x-z)) \left(a^\dagger_x a^\dagger_y a^\dagger_z - a_x a_y a_z \right),
\]
with $\omega_{\cM}$ the minimizer of~\eqref{eq:def_b}. This operator $B$ has been chosen so that, when expanding via the Duhamel formula 
\[
	e^{-B} X e^B = \sum_{k\geq 0} (-1)^{k} \ad^{(k)}_B(X) / k!\,,
\]
the cubic term $\cL_3$ is renormalized. More precisely, because the function $f_{\cM} = 1 -\omega_{\cM}$ solves~\eqref{eq:scat_M}, the operator $B$ satisfies at leading order
\begin{align}
	\left[ \d\Gamma (-\Delta) + \cL_6, B \right] + \cL_3 &\simeq 0 \label{eq:double_comm1}
	\intertext{and}
	\frac{1}{2} \left[ \left[ \d\Gamma (-\Delta) + \cL_6, B \right], B \right] + [\cL_3,B] &\simeq \frac{1}{6} n^3 \ell^{-4} \left( b_{\cM}(V) - \int_{\R^6} V \right). \label{eq:double_comm2}
\end{align}
We therefore choose the ansatz
\[
	\Psi = \cW \exp(B) | 0 \rangle\,,
\]
where $| 0 \rangle$ is the vacuum vector. This gives the upper bound matching~\eqref{eq:bound_H_nl}. To construct the ansatz on the initial box $\Omega= [-L/2,L/2]^3$, we simply patch the ground states $\Psi_z$ on the small boxes 
\[
	B_z = (\ell+R) z + [0,\ell]^3, \quad z \in \llbracket 0,M-1\rrbracket^3\,,
\]
where $M^3$ is the number of boxes and $R>0$ is some safe distance controlling the interaction between different boxes.

%%%%%%%%%%%%%%%%%%%%%%%%%%%%%%%%%%%%%%%%%%%%%%%%%%%%%%%%%
\section{The Gross--Pitaevskii regime} 
%%%%%%%%%%%%%%%%%%%%%%%%%%%%%%%%%%%%%%%%%%%%%%%%%%%%%%%%%
In this setting, we consider $N$ bosons in $\R^3$ that are trapped by a confining potential $V_{\rm ext}: \R^3 \to \R$. The system is described by the Hamiltonian
\begin{equation}\label{eq:HN}
	H_{N}^{\rm GP} = \sum_{i=1}^N\left( -\Delta_{x_i} + V_{\rm ext}(x_i)\right) + \sum_{1\le i < j < k \le N} N V\left(N^{1/2}(x_i - x_j,x_i-x_k)\right),
\end{equation}
acting on $L^2_s(\R^{3N})$. We require the external potential to be locally bounded and to grow to infinity fast enough:
\begin{equation}\label{eq:Vext}
	V_{\rm ext} \in L^{\infty}_{\rm loc}(\R^3,\R) \quad \textrm{ and } \quad V_{\rm ext}(x) \ge C |x|^\alpha\,, \quad \textrm{ for some } C>0\,, \, \alpha>0\,.
\end{equation}

As explained in the previous section, in the Gross--Pitaevskii regime the gap of the kinetic operator is of the order of the ground state energy per particle $N^2 b_{\cM}(NV(N^{1/2}\cdot)) \simeq 1$. At this length scale, boundary conditions do matter and the energy is effectively described to the leading order by the 3D energy-critical nonlinear Schr{\"o}\-dinger (NLS) functional 
\begin{equation}\label{eq:NLS}
	\cE_{\rm GP}(u)= \int_{\R^3} \left( |\nabla u(x)|^2 + V_{\rm ext}(x) |u(x)|^2 + \frac{b_{\cM}(V)}{6} |u(x)|^6 \right) \d x\,. 
\end{equation}

In~\cite{NamRicTri-21} we proved the following result.

\begin{theorem}[Ground state energy in the Gross--Pitaevskii regime]\label{thm_GS_energy_GP}
	Let $0\le V \in L^\infty(\R^6)$ be compactly supported and satisfy the three\nobreakdash-body symmetry~\eqref{eq:sym}. Let $V_{\rm ext}$ be as in~\eqref{eq:Vext}. Then the ground state energy of $H_N^{\rm GP}$ in~\eqref{eq:HN} satisfies
	\[
		\lim_{N\to \infty} \frac{\inf\sigma(H_N^{\rm GP})}{N} = e_{\rm GP} := \inf_{\|u\|_{L^2(\R^3)}=1} \cE_{\rm GP}(u)\,,
	\]
	where the effective functional $\cE_{\rm GP}$ is given by~\eqref{eq:NLS}. 
\end{theorem}

By a refinement of the proof of Theorem~\ref{thm_GS_energy_GP}, we also obtain the convergence of states for approximate minimizers.
\begin{theorem}[Condensation of approximate ground states]\label{thm_condensation_GP}
	Let $V$ and $V_{\rm ext}$ be as in Theorem~\ref{thm_GS_energy_GP}. Assume that $\Psi_N$ is an approximate ground state of $H_N^{\rm GP}$, namely
	\[
		\|\Psi_N\|_{L^2(\R^{3N})}=1 \quad \textrm{ and } \quad \lim_{N\to \infty} \frac{\left\langle \Psi_N, H^{\rm GP}_N \Psi_N\right\rangle}{ N} =e_{\rm GP}\,.
	\]
	Then we have 
	\[
		\lim_{N\to \infty} \gamma_{\Psi_N}^{(1)} = \ket{u_0}\bra{u_0}
	\]
	in trace norm, where $u_0$ is the unique nonnegative minimizer of $e_{\rm GP}$. 
\end{theorem}

Indeed, the condensation in Theorem~\ref{thm_condensation_GP} can be deduced from a standard Hellmann--Feynman argument. In this approach, the expectation of every approximate ground state $\Psi_N$ of $H_N^{\rm GP}$ against the condensate $u_0$ can be written as 
\begin{align*}
	\langle u_0, \gamma_{\Psi_N}^{(1)} u_0\rangle &= \frac{\langle \Psi_N, \sum_{i=1}^N P_i \Psi_N\rangle}{N} = \frac{\langle \Psi_N, (H_N^{\rm GP} + \eps \sum_{i=1}^N P_i) \Psi_N\rangle - \langle \Psi_N, H_N^{\rm GP} \Psi_N\rangle }{N \eps}\\
	&\ge \frac{\inf\sigma(H_N^{\rm GP} + \eps \sum_{i=1}^N P_i) - N e_{\rm GP}}{N \eps} + o(1)_{N\to\infty}\,, \qquad \forall\, \eps>0\,,
\end{align*}
where $P=|u_0\rangle \langle u_0|$. By applying the result of Theorem~\ref{thm_GS_energy_GP}, suitably adapted to the perturbed Hamiltonian $H_N^{\rm GP} + \eps \sum_{i=1}^N P_i$ with $\eps\to 0^+$ slowly when $N\to \infty$, we show that $\langle u_0, \gamma_{\Psi_N}^{(1)} u_0\rangle\to 1$, which is equivalent to the desired convergence in Theorem~\ref{thm_condensation_GP}.

Note that our results also hold for systems confined in bounded sets (which formally corresponds to the case $V_{\rm ext}=\infty$ outside the set). The same proofs work without significant modifications. In particular, for translation-invariant systems in the unit torus $[-1/2,1/2]^{3}$ (with $V_{\rm ext}=0$ inside the set), the ground state energy satisfies 
\begin{equation} \label{eq:Energy-CV-trans}
	\lim_{N\to \infty} \frac{\inf\sigma(H_N^{\rm GP})}{N} = \frac{b_{\cM}(V)}{6}
\end{equation}
by Theorem~\ref{thm_GS_energy_GP}, and the complete BEC on $u_0\equiv1$ holds by Theorem~\ref{thm_condensation_GP}. The analogue of the latter result for two\nobreakdash-body interactions was proved in~\cite{LieSei-02} (see also~\cite{LieSei-06,NamRouSei-15} for the extension to the general trapped case). 

By rescaling, we can think of having $N$ particles in $[-1/2,1/2]^{3}$ with a scaled potential $NV(N^{1/2}\cdot)$ as having $N$ particles in a box $[-N^{1/2}/2,N^{1/2}/2]^3$ with an unscaled potential $V$. In the latter interpretation, the density of the system is proportional to $N/(N^{1/2})^3=N^{-1/2}$, hence the result~\eqref{eq:Energy-CV-trans} in the Gross--Pitaevskii regime is consistent with Theorem~\ref{thm_GS_energy_low_density} in the thermodynamic limit. 

In comparison to the translation-invariant case, the general trapped case is significantly harder. In particular, the Temple inequality is no longer helpful for the energy lower bound. Instead we need to develop a new bootstrap argument where the regularization of the potential is done in several steps thanks to Dyson's lemma. In the following, we will only discuss the lower bound of Theorem~\ref{thm_GS_energy_GP} as it is the main novel part. The upper bound goes similarly as explained in Section~\ref{sec:TL} and the convergence of states follows from a standard argument as in the two\nobreakdash-body interaction case~\cite{LieSei-02,LieSei-06,NamRouSei-15}. 

%%%%%%%%%%%%%%%%%%%%%%%%%%%%%%
\subsection*{Sketch of the proof: lower bound}
%%%%%%%%%%%%%%%%%%%%%%%%%%%%%%
We follow the proof strategy of~\cite{NamRouSei-15}. That is, we aim to replace the singular potential $NV(N^{1/2}\cdot)$ by a mean-field type potential $N^{-2} b_{\cM}(V) R^{-6} U(R^{-1}\cdot)$ for $R \gg N^{-1/2}$ and apply mean-field techniques to conclude. A proof relying on division in smaller boxes, as for Theorem~\ref{thm_GS_energy_low_density} and in the spirit of~\cite{LieSeiYng-00}, should in principle also work here. However, the proof we propose should easily generalize to dealing with a magnetic field or a long range mean-field interaction.

The first step is to apply Lemma~\ref{lem:dyson_lemma} at the many-body level. For this we remove four\nobreakdash-body collisions. Let us denote for all $R>0$,
\[
	\chi_{R}(x) = \1_{\{|x|\le R\}} = 1 - \theta_{R}(x)
\]
and note that for all $1\le i \le N$,
\begin{align*}
	1 &= \prod_{\substack{j=1 \\ j\neq i}}^N (\chi_{R}(x_i-x_j) + \theta_{R}(x_i-x_j)) \\
	 &\geq \sum_{\substack{1\le j<k \le N \\ j \neq i \neq k}} \chi_{R}(x_i-x_j)\chi_{R}(x_i-x_k) \prod_{\ell \neq i,j,k} \theta_{R}\left(x_i-x_\ell \right) \\
	 &\geq \sum_{\substack{1\le j<k \le N \\ j \neq i \neq k}} \chi_R(x_i-x_j)\chi_R(x_i-x_k) \chi_R(x_j-x_k) \prod_{\ell \neq i,j,k} \theta_{2R}\left(\frac{x_i+x_j+x_k}{3} -x_\ell\right).
\end{align*}
Inserting this inside $-\Delta_{x_i}$ allows us to apply Lemma~\ref{lem:dyson_lemma} on each term and to obtain
\begin{equation} \label{eq:intro-many-body-Dyson}
\begin{multlined}[b]
	 (1-\eps)^{-1} H_N^{\rm GP} + C_\eps R^3 N^2 \\
	\!\!\!\!\!\!\!\! \ge \sum_{i=1}^N h_i + \frac{1}{6 N^2} \sum_{\substack{ 1\le i,j,k \le N \\ i\neq j \neq k \neq i } } {U_R}(x_i-x_j, x_i-x_k) \prod_{\ell \neq i,j,k} \theta_{2R}\left(\frac{x_i+x_j+x_k}{3}-x_\ell \right),
\end{multlined}
\end{equation}
where
\[
	h = \eps p^2 + p^2 \1_{\{|p|\le \eps^{-1}\}}+ V_{\rm ext}(x)\quad \text{ and } \quad U_R = R^{-6}U(R^{-1}\cdot)\,,
\]
with $\supp \, U_{R} \subset \{R/8 \leq |\bx| \leq R/4 \}$ for $R \gg N^{-1/2}$. Note that we choose to not use the whole kinetic energy in Dyson's lemma. We keep $\eps p^2$ in order to be able to apply mean-field techniques ---this is to be compared to the use of the Temple inequality in Section~\ref{sec:TL}--- as well as $p^2 \1_{\{|p|\le \eps^{-1}\}}$ in order to recover the Gross--Pitaevskii energy~\eqref{eq:NLS} in the limit $\eps \to 0$. This is partly the reason why the error term $C_\eps R^3 N^2$ appears in~\eqref{eq:intro-many-body-Dyson}.

Now, before applying mean-field techniques, we want to remove the four\nobreakdash-body collision cut-off in the r.h.s.\ of~\eqref{eq:intro-many-body-Dyson}. A simple way to do so is to use Bernouilli's inequality
\[
	\prod_{\ell \neq i,j,k} \theta_{2R}\left(\frac{x_i+x_j+x_k}{3}-x_\ell \right) \geq 1 - \sum_{\ell \neq i,j,k} \chi_{2R}\left(\frac{x_i+x_j+x_k}{3}-x_\ell \right).
\]
Using that $U_R(x,y) \le C R^{-6} \chi_{R}(x) \chi_{R}(y)$, we need to control
\begin{equation} \label{eq:4body}
	\frac{1}{N^2R^6} \sum_{\substack{ 1\le i,j,k \le N \\ i\neq j \neq k \neq i } } \sum_{\ell \neq i,j,k} \chi_{R}(x_i-x_j) \chi_{R}(x_i-x_k) \chi_{R}(x_i-x_\ell)\,.
\end{equation}
In the two\nobreakdash-body case, this term can be controlled using moment estimates and the operator inequality 
\[
	W(x-y)\le C_\eta \|W\|_{L^1(\R^3)} (1-\Delta_x)^{3/4+\eta} (1-\Delta_y)^{3/4+\eta},\quad \forall\, \eta>0\,.
\]
However, the three\nobreakdash-body generalization of this inequality, which reads
\[
	W(x-y,x-z) \le C_\eta \|W\|_{L^1(\R^6)} (1-\Delta_x)^{1+\eta} (1-\Delta_y)^{1+\eta} (1-\Delta_z)^{1+\eta}, \quad \forall\, \eta>0\,,
\]
is not enough for our purposes. Indeed, the above inequality barely fails to control the potential by the kinetic energy. This has to be compared with the embedding $H^{3+3\eta}(\R^6) \subset L^\infty(\R^6)$, which becomes false at $\eta = 0$. To circumvent this problem, following~\cite{LieSei-06} and denoting $\Psi_N$ the ground state of $H_N^{\rm GP}$, we prove that
\[
	\left\langle\Psi_N, \prod_{i=2}^4 \chi_{R}(x_1-x_i) \Psi_N \right\rangle \le C R^{9}\,
\]
holds up to a subsequence.
Hence, the expectation of~\eqref{eq:4body} is of order $N \times N R^{3}$. Since we want it to be subleading, this imposes the condition
\[
	N^{-1/2} \ll R \ll N^{-1/3}\,.
\]
Unfortunately, this constraint does not allow us to apply mean-field techniques yet, as they require $R \gg N^{-\delta}$ for some $\delta>0$ small enough. The restriction $R \ll N^{-1/3}$ allows to control the error made by discarding four\nobreakdash-body collisions ---compare with the proof of Theorem~\ref{thm_GS_energy_low_density} in Section~\ref{sec:TL} or in~\cite{NamRicTri-22}, where the effect of four\nobreakdash-body collisions is much weaker since there are fewer particles because the side length of the boxes is much smaller than the Gross--Pitaevskii length scale.

This difficulty is solved by using the bosonic symmetry of $\Psi_N$ and rewriting the energy as the expectation of a Hamiltonian with fewer particles. Namely,
\begin{multline*}
	\frac{1}{N} \left\langle \Psi_N, \Bigg( \sum_{i=1}^N h_i + \frac{1}{6N^2} \sum_{\substack{ 1\le i,j,k\le N \\ i\ne j\ne k \ne i}} U_{R} (x_i-x_j, x_i-x_k) \Bigg) \Psi_N \right \rangle \\
	\approx \frac{1}{N_1} \left\langle \Psi_N, \Bigg( \sum_{i=1}^{N_1} h_i + \frac{1}{6{N_1}^2} \sum_{\substack{ 1\le i,j,k\le N \\ i\ne j\ne k \ne i}} U_{R} (x_i-x_j, x_i-x_k) \Bigg) \Psi_N \right \rangle
\end{multline*}
for every $N \ge N_1 \gg 1$. Then repeating the previous argument, we can replace $U_{R}$ by $U_{R_1}$ provided the following condition holds
\[
	N_1^{-1/3} \gg R_1 \gg R \gg N_1^{-1/2}.
\]
Iterating this argument allows to replace the potential by a mean-field one $N^{-2} U_{R_p}$ with $R_p \gg N^{-\delta_p}$ where $\delta_p \to 0^+$ as $p \to \infty$. The rest of the proof follows a classical mean-field approximation (see e.g. the method in~\cite{LNR-16}), which we omit. 

%%%%%%%%%%%%%%%%%%%%%%%%%%%%%%%%%%%%%%%%%%%%%%%%%%%%%%%%% 
\section{Generalizations and conjectures} 
%%%%%%%%%%%%%%%%%%%%%%%%%%%%%%%%%%%%%%%%%%%%%%%%%%%%%%%%%
There are several directions to generalize our results. In this section, we discuss some open questions and formulate some conjectures for future studies. 

%%%%%%%%%%%%%%%%%%%%%%%%%%%%%%
\subsection{Second order expansion in the Gross--Pitaevskii regime}
%%%%%%%%%%%%%%%%%%%%%%%%%%%%%%
Recall that the ground state energy of $H_N^{\rm GP}$ in~\eqref{eq:HN} satisfies 
\[
	E_N = N e_{\rm GP} + o(N)\,.
\]
Contrarily to the two\nobreakdash-body interaction case where the next order contribution is $\mathcal O(1)$~\cite{BBCS-19,BreSchSch-22,NamTri-21}, we believe that in the three\nobreakdash-body interaction case, the next order should be proportional to $N^{1/2}$. Extracting exactly this contribution is an interesting open question. 

Let us explain our prediction. The condensate, which lives on a scale of order $1$, effectively creates a two-body interaction for excitations which can now interact as soon as as two of them are at distance $N^{-1/2}$. To be more precise, keeping the same notations as in Section~\ref{sec:TL} and in view of~\eqref{eq:excitation_H}, let us consider $\cL_2$. Since we look at the Gross--Pitaevskii regime, we set $n=N=\ell^2$ and the main contribution in $\cL_2$ is therefore given by
\begin{equation}\label{eq:L2}
	\cL_2 \simeq \frac{1}{2}\int N^{3} V(N^{1/2}(x-y,x-z)) u_0(x)u_0(y) u_0(z)^2 (a^\dagger_x a^\dagger_y +a_x a_y)\,,
\end{equation}
which acts as an effective two\nobreakdash-body interaction $N^{3/2}V_{\rm 2B}(N^{1/2}(x,y))$ where
\[
	V_{\rm 2B}(x,y)= \int V(x-y,z) u_0(N^{-1/2}(x+z))^2 \d z\,.
\]
Note that here, the minimizer $u_0$ of the Gross--Pitaevskii energy is not the constant function as in Section~\ref{sec:TL} where Neumann boundary conditions were considered. The term in~\eqref{eq:L2} cannot be controlled by $\d\Gamma( -\Delta + 1 )$ or $\cL_6$ and, with standard quadratic renormalization methods, it yields a contribution of order $N^{1/2}$. Another term, $\cL_4$, is believed to contribute to the order $N^{1/2}$ of the energy, let us explain why. Recall, that the leading order is obtained by a renormalization of $\cL_3$ via the conjugation of some unitary $e^{B}$ and computed using the Duhamel formula, see~\eqref{eq:double_comm1}--\eqref{eq:double_comm2}. This procedure also yields a contribution coming from $\cL_4$. For example, let us consider the term
\[
	\cL_4' := \frac{1}{2}\int N^2V(N^{1/2}(x-y,x-z)) u_0(z)^2 a^\dagger_x a^\dagger_y a_x a_y\,.
\]
We have
\begin{multline}\label{eq:L4_BB}
	\frac{1}{2} \left[ \left[\cL_4',B \right], B \right] \simeq \frac{1}{2} \int N^{5} V(N^{1/2}(x-y,x-z)) \omega(N^{1/2}(x-y,x-t))^2 \times \\
	\times u_0(x)^2u_0(y)^2u_0(z)^2u_0(t)^2 \simeq N^{1/2}\,.
\end{multline}

Note that our consideration here holds at the operator level. Namely, the energy contribution of order $N^{1/2}$ can be extracted from suitable unitary transformations. This suggests that, in principle, both upper and lower bounds can be obtained in this manner, although a rigorous upper bound could be easier to see thanks to a trial state argument.
We therefore expect the following.

\begin{conjecture}[Ground state energy]
	We have
	\[
		E_{\rm GP}(N) = N e_{\rm GP} + N^{1/2} C^{(2)}_{\rm GP} + \mathcal O(1)\,,
	\]
	for some constant $C^{(2)}_{\rm GP}$ independent of $N$. 
\end{conjecture}

In~\cite{NamRicTri-21}, we showed that the grand canonical energy (the infimum over the whole Fock space) is bounded from above by $N e_{\rm GP} + CN^{1/2}$, and deduced the canonical upper bound $E_{\rm GP}(N) \le N e_{\rm GP} + C N^{2/3}$. In fact, our method can be refined to match the canonical upper bound $E_{\rm GP}(N) \le N e_{\rm GP} + C N^{1/2}$, and hence the sharp upper bound seems reachable, although more work should be done to capture the constant $C^{(2)}_{\rm GP}$. The main question is to get the matching lower bound. 

%%%%%%%%%%%%%%%%%%%%%%%%%%%%%%
\subsection{Excitation spectrum in the Gross--Pitaevskii regime}
%%%%%%%%%%%%%%%%%%%%%%%%%%%%%%
Although the second order of the ground state energy of $H_N^{\rm GP}$ in~\eqref{eq:HN} is believed to be proportional to $\mathcal{O}(N^{1/2})$, we expect that the excitation spectrum is still of order $\mathcal{O}(1)$ and should be described by Bogoliubov's theory, similarly to the two\nobreakdash-body interaction case~\cite{BBCS-19,BreSchSch-22,NamTri-21}. 

Taking the viewpoint in~\cite{LNSS-15}, we can predict the excitation spectrum by quantizing the Hessian of the Gross--Pitaevskii functional $\cE_{\rm GP}$. More precisely, for $\varphi \perp u_0$ we have
\[
	\cE_{\rm GP}\left(\frac{u_0 + \varphi}{\sqrt{1 + \|\varphi\|_{L^2(\R^3)}^2}}\right) = e_{\rm GP} + 
	\frac{1}{2}
	\left\langle
		\begin{pmatrix} \varphi \\ \varphi \end{pmatrix}, \cE_{\rm GP}''(u_0) \begin{pmatrix}\varphi \\ \varphi \end{pmatrix}
	\right\rangle
	+ o\!\left(\|\varphi\|_{H^1(\R^3)}^2\right),
\]
with the Hessian matrix
\[
	\cE_{\rm GP}''(u_0) =
	\begin{pmatrix}
	D + \frac{1}{2} b_{\cM}(V) u_0^4 & \frac{1}{2} b_{\cM}(V) u_0^4 \\
	\frac{1}{2} b_{\cM}(V) u_0^4 & D+\frac{1}{2} b_{\cM}(V) u_0^4
	\end{pmatrix}
\]
and where 
\[
	D = -\Delta + V_{\rm ext} + \frac{b_{\cM}(V)}{2} u_0^4 - \mu \quad \textrm{ and } \quad \mu = \int_{\R^3} |\nabla u_0|^2 + \frac{b_{\cM}(V)}{2} \int_{\R^3} u_0^6 \,.
\]
Diagonalizing the Hessian by a real symplectic matrix (see~\cite{NamTri-21} for details), we can find the excitation operator $E = (D^{1/2}\left(D + b_{\cM}(V) u_0^4\right)D^{1/2})^{1/2}$. Hence, we formulate the following conjecture.
\begin{conjecture}[Excitation spectrum]
	The low-lying spectrum of $H_N^{\rm GP}$ consists of finite sums of the form
	\[
		\inf \sigma (H_N^{\rm GP}) + \sum_{i\geq 1} n_i e_i\,,
	\]
	where $n_i \in \{0,1,2,\dots\}$ and $\{e_i\}_{i=1}^\infty$ are the positive eigenvalues of the operator
	\[
		E = \left(D^{1/2}\left(D + b_{\cM}(V) u_0^4\right)D^{1/2}\right)^{1/2}.
	\]
\end{conjecture}

%%%%%%%%%%%%%%%%%%%%%%%%%%%%%%
\subsection{Combined two- and three-body interactions}
%%%%%%%%%%%%%%%%%%%%%%%%%%%%%%
So far, for simplicity, we have considered systems with only three\nobreakdash-body interactions. However, a more realistic model than~\eqref{eq:HN} would also take into account two\nobreakdash-body interactions. Let us denote $W: \R^3 \to [0,\infty)$ an even, bounded and compactly supported potential, and consider 
\begin{multline}\label{eq:HN_comb}
	H_{N}^{\rm Comb} = \sum_{i=1}^N\left( -\Delta_{x_i} + V_{\rm ext}(x_i)\right) + \sum_{1\le i < j \le N} N^{3\beta -1} W\!\left(N^\beta(x_i-x_j)\right) \\
	+ \sum_{1\le i < j < k \le N} N V\!\left(N^{1/2}(x_i - x_j,x_i-x_k)\right),
\end{multline}
where $\beta \in (0,1]$ is a parameter to adjust the range of two\nobreakdash-body interactions. In this case, we expect for the leading order of the ground state energy of $H_N^{\rm Comb}$ to be given by the Gross--Pitaevskii energy 
\[
	e^{\rm Comb}_{\rm GP}(b_1,b_2) = \inf_{\|\varphi\|_{L^2(\R^3)} = 1} \cE^{\rm Comb}_{\rm GP}(b_1,b_2)(u)
\]
where 
\begin{equation}\label{eq:E-comb}
	\cE^{\rm Comb}_{\rm GP}(b_1,b_2)(u) := \int_{\R^3} \left(|\nabla u|^2 + V_{\rm ext} |u|^2\right) + \frac{b_1}{2} \int_{\R^3} |u|^4 + \frac{b_2}{6} \int_{\R^3} |u|^{6}\,,
\end{equation}
for suitable constants $b_1$ and $b_2$.

For $\beta < 1$, an adaptation of our proof in~\cite{NamRicTri-21} is expected to yield 
\begin{equation}
	\lim_{N\to \infty} \frac{\inf \sigma (H_N^{\rm Comb})}{N} = e^{\rm Comb}_{\rm GP}\left(\|W\|_{L^1(\R^3)}, b_{\cM}(V)\right).
\end{equation}
The case $\beta = 1$ is more challenging as it corresponds to the situation where each of the interaction potentials is in its critical scaling, and hence the correlations at each of the two- and the three\nobreakdash-body levels contribute to the leading order.

\begin{conjecture}[Ground state energy]
	Taking $\beta = 1$ in~\eqref{eq:HN_comb}, we have
	\[
		\lim_{N\to \infty} \frac{\inf \sigma (H_N^{\rm Comb})}{N} = e^{\rm Comb}_{\rm GP}\left(b(W), b_{\cM}(V)\right),
	\]
	with $b(W)$ the scattering energy defined in~\eqref{eq:def-scat-energy} and $b_{\cM}(V)$ defined in~\eqref{eq:def_b}. 
\end{conjecture}

Recall that when only one of the two interactions is there, namely either $V=0$ or $W=0$, we may employ a Dyson's lemma like Lemma~\ref{lem:dyson_lemma} to replace the singular potential by a mean-field type one, at least for the energy lower bound. However, in the case of combined interactions, it is not clear how to apply the lemma to both potentials. This would require a clever separation of scales such that each correlation process should not interfere with the other.

Another direction to investigate is the case where $V\ge 0$ but $W\le 0$. Interestingly, the physics literature suggests that the competition between repulsive and attractive interactions can lead to a so-called droplet state~\cite{Petrov-14}, where the condensate is self-trapped (without external field), and even to crystalline structures~\cite{BB-15,B-16}.

At the level of the Gross--Pitaevskii functional~\eqref{eq:E-comb}, by interpolation of the $L^4$ norm between the $L^2$ and $L^6$ norms, we have $e^{\rm Comb}_{\rm GP}(b_1,b_2) > -\infty$ no matter the sign of $b_1$ as long as $b_2>0$. We refer to~\cite{LN-20} for a detailed analysis of the effective equation. 

At the many-body level, however, it remains unclear even which conditions are needed on $\beta$ for the stability of the second kind to hold:
\[
	\exists\, C>0\,, \quad H_{N}^{\rm Comb} \geq - C N\,.
\]
We expect that this stability holds if $\beta>0$ is small, and possibly up to $\beta=1/2$ where the two- and three\nobreakdash-body interactions have the same length scales. In contrast, when $\beta>1/2$, the short-length scale of the two\nobreakdash-body interactions may lead to a severe instability.

%%%%%%%%%%%%%%%%%%%%%%%%%%%%%%
\subsection{Second order in the thermodynamic limit}
%%%%%%%%%%%%%%%%%%%%%%%%%%%%%%
In the thermodynamic limit, the ground state energy per unit volume of dilute Bose gases with two\nobreakdash-body interactions was predicted in the 1950s~\cite{LeeHuaYan-57,Wu-59} to be given by 
\begin{equation}\label{eq:e_rho}
	e_{\rm 2B}(\rho) = 4\pi a \rho^2 \left(1 + \frac{128}{15\sqrt \pi} \sqrt{\rho a^3} + 8\left(\frac{4\pi}{3} - \sqrt 3\right) \rho a^3 \log\!\left(\rho a^3\right)+ \mathcal O\!\left(\rho a^3\right)\right)
\end{equation}
as $\rho a^3 \to 0$, where $\rho$ is the density and $a$ is the scattering length of the interaction potential. As we mentioned earlier, the first order has been established by Dyson~\cite{Dyson-57} (upper bound) and Lieb--Yngvason~\cite{LieYng-98} (lower bound). The second order term in~\eqref{eq:e_rho} is much more involved. This so-called Lee--Huang--Yang correction is heuristically obtained in two steps: diagonalizing the quadratic contributions ---Bogoliubov's approximation---, then replacing $\|W\|_{L^1}$ by $b(W)$ ---Landau's correction--- accounting for the correlation induced by the cubic and quartic terms neglected in the first step. The lower bound for this second order term was settled very recently by Fournais--Solovej~\cite{FouSol-20,FouSol-21}, while the upper bound was already proved in 2009 by Yau--Yin~\cite{YauYin-09} (see also~\cite{BCS-21} for a new proof of the upper bound). Understanding the third order term in~\eqref{eq:e_rho} from first principles remains very challenging.

Now, concerning the ground state energy of dilute Bose gases with three\nobreakdash-body interactions, two questions naturally arise :

\textbf{1.} Does $e_{\rm 3B}(\rho)$ satisfy a similar expansion, in the dilute regime $\rho b_{\cM}(V)^{3/4} \to 0$, to the one in the two\nobreakdash-body case~\eqref{eq:e_rho}?

\textbf{2.} Does this expansion show universality? That is, do next orders only depend on $\rho$ and $b_{\cM}(V)$?

To answer these questions, let us first carry out a similar heuristic argument in the Gross--Pitaevskii regime. More precisely, let us consider again~\eqref{eq:excitation_H}, but looking now at the grand-canonical case for simplicity. The two candidates for the next to leading order term are the contributions coming from the quadratic terms and the ones coming from the quartic terms. Recall that, in the Gross--Pitaevskii regime, the renormalization of $\cL_3$ ---see~\eqref{eq:L3}--- is responsible for the appearance of a term of order $N^{1/2}$ due to the presence of the quartic terms. Let us assess this contribution in the thermodynamic limit. In view of~\eqref{eq:excitation_H}, taking $n=N$ and $\ell = L$, we find, similarly as in~\eqref{eq:L4_BB}, 
\begin{multline*}
	\frac{1}{2} \left[ \left[ \cL_4',B \right], B \right] = \frac{1}{2} \int N L^2 V(L (x-y,x-z)) N^3 \omega(L(x-y,x-t))^2 \times \\
	\times u_0(x)^2u_0(y)^2u_0(z)^2u_0(t)^2 \simeq \rho^4 b_{\cM}(V)^{7/4} L^5\,.
\end{multline*}
This can be seen, for example, by denoting $V = b_{\cM}(V)^{-1/2} \widetilde V(b_{\cM}(V)^{-1/4}\cdot)$ where the potential $\widetilde V$ has a fixed scattering energy $b_{\cM}(\widetilde V) = 1$. Now, recalling that $ H_{N,L} = L^{-2} \mathcal U^* \widetilde{H}_{N,L} \mathcal U$, the contribution per unit volume of the quartic term is $\rho^4 b_{\cM}(V)^{7/4}$.

Similarly, the quadratic terms give a contribution of the same order than the quartic ones. Indeed, in the thermodynamic limit, the term~\eqref{eq:L2} becomes
\[
	\cL_2 \simeq \frac{1}{2}\int N^{2} L^2 V(L(x-y,x-z)) u_0(x)u_0(y) (a^\dagger_x a^\dagger_y +a_x a_y)\,.
\]
From standard analysis on quadratic operators, we know that
\[
	\d\Gamma(-\Delta) + \cL_{2} \gtrsim - \frac{1}{2} \tr K_{2B} (-\Delta)^{-1/2} K_{2B} \simeq - C \rho^4 b_{\cM}(V)^{7/4} L^5\,,
\]
where $K_{2B}$ is the operator with kernel 
\[
	K_{2B}(x,y) = \int N L^2 V(L(x-y,x-z)) u_0^2(z) \d z\,.
\]

Taking the above expression to the infinite volume limit, we arrive at the following.

\begin{conjecture}[Ground state energy]
	In the thermodynadmic limit, for fixed $V$, the ground state energy in the low density regime $\rho \to 0$ satisfies 
	\[
		e_{\rm 3B}(\rho) = \frac{1}{6} b_{\cM}(V) \rho^3 \left(1 + C^{(2)}_{\rm TL}(V) \rho + o(\rho) \right)
	\]
	with a constant $C^{(2)}_{\rm TL} (V) \in \R$.
\end{conjecture}
It is unclear to us whether $C_{\rm TL}^{(2)}(V)$ depends on V only via the scattering energy $b_{\cM}(V)$ or not.

\bigskip
\raggedbottom

\end{document}